# Applying a Text-Based Affective Dialogue System in Psychological Research:
## Case Studies on the Effects of System Behavior, Interaction Context and Social Exclusion

Marcin Skowron

*Austrian Research Institute for Artificial Intelligence*
*Freyung 6/6/3a, Vienna, Austria*

marcin.skowron@ofai.at

Stefan Rank

*Drexel University, Westphal College of Media Arts & Design*
*3141 Chestnut Street, Philadelphia, USA*

stefan.rank@drexel.edu

Aleksandra Świderska

Dennis Küster

Arvid Kappas

*Jacobs University, School of Humanities & Social Sciences - SHSS*

*Campus Ring 1, 28759 Bremen, Germany*

a.swiderska,d.kuester,a.kappas@jacobs-university.de

Abstract

This article presents two studies conducted with an affective dialogue system in which text-based system-user communication was used to model, generate, and present different affective and social interaction scenarios. We specifically investigated the influence of *interaction context and roles* assigned to the system and the participants, as well as the impact of *pre-structured social interaction patterns* that were modelled to mimic aspects of "social exclusion" scenarios. The results of the first study demonstrate that both the social context of the interaction and the roles assigned to the system influence the system evaluation, interaction patterns, textual expressions of affective states, as well as emotional self-reports. The results observed for the second study show the system's ability to partially exclude a participant from a triadic conversation without triggering significantly different affective reactions or a more negative system evaluation. The experimental evidence provides insights on the perception, modelling and generation of affective and social cues in artificial systems that can be realized in different modalities, including the text modality, thus delivering valuable input for applying affective dialogue systems as tools for studying affect and social aspects in online communication.

# 1 Introduction

The exploration of artificial systems that are able to recognize, process and model social context in complex human-human and human-computer interactions poses highly interdisciplinary research problems. Designers of such systems draw from research on how humans perceive, react to and interact with each other, but numerous challenges remain: A more complete understanding of underlying natural phenomena, robust modelling of social context and the realization of corresponding functionality in computer systems, as well as the comprehensive evaluation of their influence on humans. In recent years, many aspects that are crucial for these advances have been addressed, but the discussion of appropriate design principles and suitable computational tools is far from complete.

The fields of affective computing, social signal processing and human-computer interaction (HCI), incorporate a wide range of methods for the detection and interpretation of affect (Calvo and D'Mello, 2010) and social signals (Vinciarelli et al, 2012), along with different models implementing selected aspects of affective and social processes in interactive systems. This line of research also abounds in studies on multimodal communication as well, related especially to the perception, modelling, and generation of non-verbal behaviour. The investigation of the corresponding methods in a predominantly text-based modality received comparatively little attention. Another motivation for research in this area is the unprecedented level of users' communication in online social media, frequently realized in a text-based form. The resultant growing scope of application for socially intelligent autonomous agents requires abilities to detect, interpret, engage in, or conduct social interaction scenarios relying on various modalities characteristic for and available in different environments. In particular, we consider autonomous text-based dialogue systems that combine the modelling of selected aspects of affective and social processes in interaction scenarios, assisting the systematic studies of their effects on users (Skowron et al, 2013).

Previous work conducted with the dialogue system applied in the current studies - "Affect Listener" - presented an integrated view on a series of experiments, in which the system modelled selected aspects of *affective and social processes in online, text-based communication*. Specifically, three different realizations of the system were evaluated to assess the effects of affective profiles and fine-grained communication scenarios on users' expressions of affective states, experienced emotional changes, and interaction patterns. The obtained results demonstrated that the system applied in virtual reality settings matched a Wizard-of-Oz in terms of chatting enjoyment, dialogue coherence and realism in short interactions. The realized variants of the system's affective profile (Skowron et al, 2011b) (positive, negative and neutral) significantly influenced the rating of chatting enjoyment and an emotional connection. Self-reported emotional changes experienced by participants during the interaction with the system were in line with the type of applied profile, e.g., a positive affective profile elicited positive emotional changes in users. Analysis of interaction logs, including the usage of particular dialogue acts, word categories, and textual expressions of affective states for the realized fine-grained scenarios, demonstrated the system's ability to successfully enable a scenario of "social sharing of emotions".

This paper presents two new studies recently conducted with the "Affect Listener" in which the system-user communication, limited to text modality, was used to perceive, model and generate different social interaction characteristics. In particular, we investigated the role of *interaction context and roles* assigned to the system and the participants as well as the influence of *structuring of social interactions*, in this case, the realization of a *social exclusion scenario*, on the communication style of users, textual expressions of affect, self-reported emotional states, and impression formation regarding conversational partners. We also analysed how these components relate to each other and account for the changes introduced artificially or occurring naturally during the course of online interactions. This includes the application of sentiment mining and affect analysis tools and resources, and their integration with dialogue management and generation components.

The next section presents an overview of relevant background. The following sections introduce the method applied in the studies, i.e., an overview of the affective dialogue system, and a description of the interaction scenarios used in the experiments. We then discuss the experimental results from both studies and conclude by providing a summary of the main findings and the contribution of the current work.

## 2 Relevant Work

Both of the studies presented in this article explored the intriguing intersection between computer science, artificial intelligence and social psychology. While our main contribution pertains to the technical advances related to the use and implementation of artificial conversation systems as discussed above, the emerging capabilities of these types of systems are becoming increasingly relevant also for psychologists in the study of social phenomena. At a very general level, these psychological research interests may be the consequence of a shift of attention onto the impact of computer-mediated communication - a discussion that is still ongoing. For example, due to the explosion of research on specific environments of online communication, such as Facebook, the apparent lack of direct human to human social connection and the resulting consequences of loneliness associated with the increasing use of online communication have received a lot of critical attention (see e.g., Cacioppo, 2009). The question arises, to what extent we are able to connect psychologically with an interaction partner through text and how this is limited. Hence, the use of artificial systems as interaction partners opens new avenues for psychological research in a variety of domains, for example social exclusion paradigms (e.g., Kassner, Wesselmann, Law, & Williams, 2012; Williams, Yeager, Cheung, & Choi, 2012), economic games (e.g., the Ultimatum Game, Sanfey et al., 2003), or objectification, dehumanization, and agency (Ward et al., 2013). The two studies presented in this paper relied on two distinct but related paradigms to address the more general question of to what extent artificial conversation systems were perceived and responded to as if they were a human conversation partner and how the interactions with these systems diverged from human-human interaction.

Prior research has demonstrated that humans have a natural tendency to attribute human qualities to non-living entities (Waytz, Epley, & Cacioppo, 2010). For example, it has been found that people readily attribute the power to act or to intentionally behave in a certain way to animals, gods, and technological gadgets (Epley, Akalis, Waytz, & Cacioppo, 2008; Waytz et al., 2010). One of the most scrutinized inanimate social actors has been the computer (e.g., Reeves & Nass, 1996). In this context, experiments that investigated the mechanisms of human-human interactions (HHI) were systematically adapted to the study of human-computer interactions (HCI). That is, participants' interaction partners were substituted with a regular computer. It has been then shown that, for example, people adhere to politeness norms when they reply to computers (Nass, Moon, & Carney, 1999), apply gender stereotypes (Nass, Moon, & Green, 1997), and take part in mutual self-disclosure (Moon, 2006). Some authors argued that the computer turned out to temporarily become a fellow human to the participants and the virtual world to become real (Reeves & Nass, 1996). The term "media equation" was coined to refer to these effects ("media equals real life"; Reeves & Nass, 1996).

It has been further established that in HCI, people accurately recognize and rapidly assess types of personality conveyed by the interactive artificial systems, even when the available cues are minimal (Nass, Moon, Fogg, Reeves, & Dryer, 1995). They also react to the computers' personalities analogously to how they would typically approach humans exhibiting such qualities (Isbister & Nass, 2000; Moon & Nass, 1996). Moreover, in line with the similarity-attraction hypothesis, people tend to prefer individuals who match their own traits (e.g., Blankenship, Hnat, Hess, & Brown, 1984, in Nass et al., 2000) and this has been shown to be the case also for computers that appear to be similar to the user (Nass et al., 2000, 2005). Specifically, the use of words, interactivity, and assignment of roles usually fulfilled by humans amounts to a set of social cues sufficient to make users treat computers as social partners and apply social rules and expectations to the interaction (e.g., Nass & Moon, 2000). All of the listed cues are present for the artificial conversational systems.

Despite the increasing level of realism and similarity to a human interaction partner that can now be achieved by artificial conversational systems, there is certainly still a gap between the capabilities of the state of the art AI and typical human interaction partners. Thus, while prior research has shown that for example, two human research confederates can be used to exclude a third human user in a chat-based social exclusion paradigm (Williams, Govan, Croker, Tynan, Cruickshank, & Lam, 2002), it remains unclear to what extent one or more artificial conversational systems would be able to convincingly perform such a complex task in place of the human confederates. Further, if only one conversational system is used to interact with two human users, the additional question arises if the perceived social connection between two acquainted human participants (see Waytz & Epley, 2012) could be used to study processes of dehumanization without the ethical implications associated with having to exclude a human participant in this type of research on social exclusion. As has been shown in related research from social psychology, participants can react strongly to social exclusion when they believe they are excluded by two other human participants, e.g., in the virtual ball-tossing game "Cyberball" (Williams, Cheung, & Choi, 2000), or in immersive virtual environments (Kassner, Wesselmann, Law, & Williams, 2012). This, however, could change if participants are aware that they will be interacting with an artificial system, or if the system is not perceived as a social actor with sufficient agency to include or exclude a human participant.

Two divergent predictions can be made regarding perceptions and behaviour of people interacting with artificial conversational systems. Following the Ethopoeia approach (Nass & Moon, 2000; von der Pütten et al., 2010), if agents or avatars exhibit enough social cues, they will elicit comparable social responses. On the contrary, according to the Threshold Model of Social Influence (Blascovich, 2002; Blascovich et al., 2002), conversational agents that are known to be artificial will be below the threshold of effective social influence, and therefore, they will not elicit the same quality of responses as a human interaction partner would in an equivalent situation. While some of the research arguing for a high degree of similarity in how humans and artificial systems are perceived has already been presented, there is likewise some initial evidence for systematic qualitative differences in the way that humans respond to another human vs. a known artificial entity. An interesting line of research that accommodated the versatile comparison of reactions to human interaction partners on the one hand and to regular computers as well as computer-driven entities (agents, avatars) on the other was founded on social decision making paradigms. For example, based on people's behaviour in bargaining games (e.g., the Ultimatum Game; Sanfey et al., 2003), it has been established that the full-fledged behavioural and emotional reactions take place only when participants believe to be playing against another person. The magnitude of negative reactions to computers is considerably smaller (Sanfey et al., 2003). It can be therefore assumed that similar effects will occur also in interactions with artificial conversational systems, for instance when the systems will be participants' partners in bargaining games or when they will be trying to exclude participants from a conversation. Alternatively, the human-like quality of engaging in a flexible dialogue with an advanced AI may already be so convincing that the systems will be treated as if they are no different from a human conversation partner. The current experiments were thus designed to measure detailed aspects of the systems' evaluations as well as the outcomes of the interactions across two different experimental paradigms, linked to extant research from social psychology. They can serve as case studies for the usefulness of text-based affective dialogue systems for this type of research and, in doing so, also underscore the maturity of the system described in this contribution.

The hypothesis of a complex interconnection between emotions and cognition

(Kappas, 2006) motivates the incorporation of affective processing in computer systems. For example, in the application domains, such as, e.g., Intelligent Tutoring Systems the interplay between motivation, affect, cognition and learning outcomes, has been identified as particularly important for improving the quality of such systems (Arnold, 1999). Also in the other interactive applications such as Affective Dialogue Systems (ADS) (Andre et al. 2004), Companion Systems (Willks, 2010), Multimodal Artificial Listeners (Schröder et al, 2012), and Text-Based Affect Listeners (Skowron et al. 2013), the integration of affective processing at the level of perception, modelling, and generation is hypothesized to improve the quality of interactions, supporting also systematic studies of affective, cognitive and social aspects in HCI. The potential effects of such interactions and their influence on a dynamics in large scale online communities was recently investigated using agent-based simulations, and a range of interactive and affective parameters obtained from experiments with different realizations of the text-based ADS (Tadic et al. 2014).

The management of human-computer conversations that account for the exchanged emotional cues is a central area of interest in the design of ADS. This multidisciplinary research field combines work on speech recognition, dialogue processing, computer graphics, animation, speech synthesis, embodied conversational agents, and human-computer interaction (Andre et al, 2004; Pittermann et al, 2009), and reflects the growing interest in emotion in human-machine interaction and as part of cognitive architectures (Ziemke and Lowe, 2009; Gros, 2010; Squartini et al., 2012). Numerous works originating from this line of research provided significant evidence that emotional factors play a pivotal role in HCI. For example, the integration of affective components was used to enhance ECA believability in tutoring systems through considering the motivational states of students, thus supporting the learning process (Bhatt et al, 2004).

More recently, different models that implement selected aspects of affective and social processes in interactive systems were presented. Their focus is often on multimodal communication, including perception, modelling and generation of non-verbal behaviour: e.g., the modelling of comprehensive listening behaviour in a virtual human architecture (Wang et al, 2011; Traum et al, 2012), constructing common grounded symbols between interlocutors and providing incremental multimodal feedback (Buschmeier and Kopp, 2013; Schröder et al, 2011), and investigating the turn-taking strategies on impression formation in Embodied Conversational Agents (ter Maat et al, 2011). The development of interactive systems, aimed at the creation of personalized conversational, multimedia interface companion systems (Willks , 2010) that are able to form long-term relationships with their users, requires the ability to detect, model and generate different conversational strategies, interactive characteristics, and their specific facets, for instance humour (Nijholt, 2007), empathy (Paiva et al. 2005), and politeness (Andre et al. 2004).

Affect sensing from text and its application to text-based dialogue systems, in particular for modelling different affective and social characteristics and studying their effects on users received comparatively less attention. For example, Tatai and Laufer (2004) presented a design for a chatterbot, which aims at providing emotionally adequate responses to user utterances and employs a keyword-based approach for the detection of emotions.

The ability to generate and convincingly convey affective communication characteristics with a text-based dialogue system was presented by Skowron and colleagues (Skowron et al, 2011a; Skowron et al. 2011b). The evaluation results showed that textual expressions of affective states by participants and self-reported emotional changes experienced when interacting with different variants of the system (i.e., positive, negative, neutral) were in line with the applied affective profile. The analysis of interaction logs and the evaluation results for the system's generated fine-grained interaction characteristics confirmed its ability to simulate scenarios such as "getting acquainted with someone" and "social sharing of emotions", and further to elicit corresponding interactive characteristics from participants (Skowron et al. 2013). The focus of this line of studies lies on the influence of structuring of emotional interactions and their systematic examination. The evaluation of the influence of the interaction context, in particular the effect of roles assigned to the system and user, and specific social interaction scenarios such as "social exclusion" was not previously studied in such an experimental setup. Thus, a goal for the studies presented here is to address these lacunae.

## 3 Method

The Affective Dialogue System (Skowron, 2010a) realizes a range of interaction scenarios, e.g., by simulating different affective profiles or following specific fine-grained communication and interaction scenarios aimed at eliciting particular affective and social processes from users, in a settings typical for online communication (Skowron et al, 2013). Previous experiments provided the following insights related to the system's functionalities, their influence on and evaluation by participants:

- Study 1 - conducted in a Wizard-of-OZ (WOZ) setting[1], validated the system's ability, on par with a human operator regarding realistic and enjoyable dialogue as well as for establishing an emotional connection with users in a short interaction. The experiment included two, 5-minutes long interactions during which simulation of thinking and typing delays was used (Skowron et al, 2011a).

- Study 2 - demonstrated the system's ability to conduct interactions where system's affective profiles, i.e., positive, negative, neutral (Skowron et al, 2011b) was consistently conveyed to the users. Further the experiments showed a significant influence of the system's affective profile on users' perception of a conversational partner, the changes of emotional states reported by the users, and on the textual expressions of affective states[2].

- Study 3 - demonstrated a successful application of the dialogue system for eliciting social sharing of emotion (Skowron et al, 2013). The experimental results also showed the impact of the realized interactions scenarios, i.e. "getting acquainted with someone" and "social sharing of emotions", on the communication style of users and on expressions of affective states, without affecting the overall perception of the conversational system.

---

[1] Participants believe that they communicate with a dialogue system, while responses are actually provided by a human operator. In the presented experiments, the operator was asked to conduct a realistic and coherent dialogue and provided free text input to user utterances.
[2] Studies 2 and 3 included three 7-minutes long interactions. No simulation of the thinking and typing delays was used. In both experiments participants were aware that they interact with an artificial system.

In summary both the evaluation results and the analysis of the participants' communication style and expressions of affective states obtained from the previous rounds of experiments support the thesis that the affective dialogue system, even restricted to the text modality, can convincingly simulate affective profiles and realize different social communication scenarios in short interactions. Further, the conducted experiments also provided strong evidence on the effects of affective and social processes simulated in the system on users' communication style and textual expressions of affective states, even when participants were aware that they interact with an artificial system.

In the applied method, the acquired data are analysed to improve our understanding of the role of emotions and social processes in online interaction. This approach relates to research on correlations between expressions of affect in text and physiological responses (Kappas et al, 2010) as well as to studies on the relations of textual communication style and content to personality traits of users (Pennebaker and King, 1999; Fast and Funder, 2008; Yarkoni, 2010).

In the previous rounds of experiments, the system was presented to the user in a virtual-bartender scenario. In Study 1, this setting was also represented graphically by a 3D bar environment (Gobron et al, 2011). The choice of setting was intended to support interaction scenarios where direct communication with a participant can be established easily. Users could relate existing knowledge regarding the behavioural norms in such a setting in the real world to the virtual environment. Furthermore, this setting also provided the flexibility to switch between open-domain chats and closed-domain dialogues of various levels of intimacy.

Continuing this line of research, we applied the text-based dialogue system in two new experiments to:

- Study 4 - assess the impact of changes introduced to the interaction setup: i.e., radical shortening of the interaction time, and different roles assigned to the system and participants ("virtual bartender" and "virtual customer" vs. "two online strangers") on users' perception of the system, especially in the aspects related to its affective profile.

- Study 5 - evaluate system's ability to realize different social interaction scenarios in text-based communication with two concurrent users, and to measure the effects of the conducted scenarios on the participants.

In the following, we present general characteristics of the system. For the detailed description of the system and its core components refer to (Skowron, 2010a; Skowron et al, 2013).

## 3.1 Affective Dialogue System

The variants of the Affect Listener system applied in the presented studies communicate with users in a text modality and use the integrated sentiment analysis and affect detection components to recognize and categorize expressions of affective states. The acquired information aids the selection and generation of responses. The system can interact with users via a range of communication channels and interfaces that share common characteristics of online chatting.

The different realizations of the system which were applied in five rounds of studies were developed based on the same software framework and share the same set of natural language processing tools and resources. The main difference between the system realizations used in studies relates to their abilities to simulate distinct affective profiles or conduct fine-grained social communication and interaction scenarios. All necessary modifications are implemented in the system's Control Layer. These changes affect, in particular, the ways in which different system realizations conduct task-oriented parts of the dialogues, e.g., opening or closing of the interaction. Further, they influence the system's responses depending on the applied scenario (e.g., the target of communication or affective behaviour set for a given system realization) and the affective states detected in user utterances. The applied method either suppresses potential responses to a detected affective state or responds in a specific way, e.g. friendly or unfriendly.

The core tasks of the system in the context of the experimental settings include: perception and classification of affective cues in user utterances and system response candidates (text-based affect detection), the incorporation of affective cues into the dialogue management and the maintenance of an emotional connection with users (affective dialogue management), management of task-oriented dialogues (closed-domain dialogue), as well as handling conversations that are not restricted in topic (open-domain chats), and, finally the detection of cues in the system-user interactions that enable the selection of (i) suitable system response generation methods (balancing task oriented dialogue vs. open-domain conversations), and (ii) an interaction partner in multi-user environments.

In the following, we introduce the software framework and layers used in the affective dialogue system; See Fig. 1 for an overview of the system architecture. The changes introduced in the Control Layer necessary for the construction of system realizations applied in the Studies 4 and 5 are summarized at the end of this section.

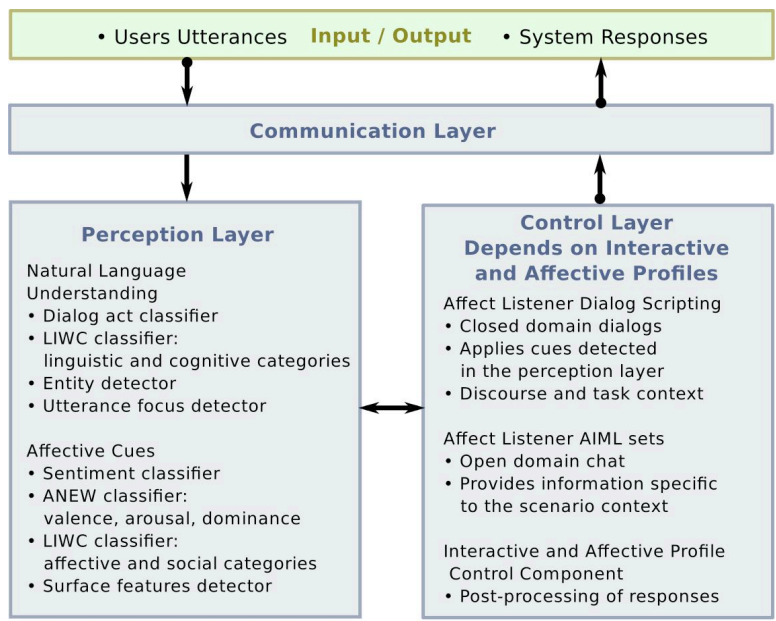

Figure 1: Layers and main components of the Affective Dialogue System.

*Perception Layer*

The Perception Layer consists of and integrates a set of natural-language processing tools, linguistic and affective resources to analyse user utterances and system response candidates:

- Dialogue Act classifier: (Skowron et al, 2011c) adaptation of the taxonomy used in the NPS Chat corpus (Forsyth and Martell, 2007). For the present scenario, the original taxonomy (Accept, Bye, Clarify, Continuer, Emotion, Emphasis, Greet, No Answer, Other, Reject, Statement, Wh-Question, Yes Answer, Yes/No Question) was extended with an additional class "Order" (food or drinks) using 339 additional training instances. For this taxonomy and training set, the SVM based DA classifier using a bag-of-words and bag-of-bigrams feature set achieved 10-fold cross validation accuracy of 76.1%. The classifier was implemented using LIBSVM (Chang and Lee, 2011) with the following settings: SVM type – C-SVM, kernel – Radial Basis Function (RBF), cost – 8.0, gamma – 0.03125.

- Lexicon-Based Sentiment Classifier: (Paltoglou et al, 2010) provides information on sentiment class, positive and negative sentiment values. It works in a rule-based manner and relies on two complementary emotional dictionaries: General Inquirer (Stone et al, 1966) and LIWC (Pennebaker et al, 2001) to obtained scores for positive and negative sentiment. The initially obtained scores are then modified with added, linguistic driven functionalities such as negation (e.g. "happy" vs. "not happy"), detection of capitalization ("bad" vs. "BAD"), exclamation and emoticons, intensifiers and diminishers to produce the final score.

- ANEW classifier (Paltoglou et al, 2013): provides information on the valence, arousal and dominance of an utterance based on the dictionary of Affective Norms for English Words (Bradley and Lang, 1999).

- LIWC (Linguistic Inquiry and Word Count) classifier (Pennebaker et al, 2001): provides linguistic, cognitive and affective categories for words. Specifically, this resource includes 32 word categories that are tapping psychological processes (e.g., affective such as positive and negative emotions; social such as family, friends and human; cognitive such as insight, causation, tentative), 22 linguistic categories (e.g., adverbs, negations, swear words), 7 personal concern categories (e.g., home, religion, work, leisure) 3 paralinguistic dimensions (fillers, assents, nonfluencies), for almost 4500 words and word stems.

- Utterance focus detector (Skowron et al, 2008); Surface features detector: e.g., exclamation marks, emoticons; Entity Detector: Gazetteers and regular expressions specific to a bar context, e.g., drinks and snacks.

The layer is responsible for the detection of affect and a range of other conversational cues that can be used to select suitable mechanisms for the generation of system response candidates and the selection of a system response in the Control Layer. The specific set of components was selected to provide the system with the necessary cues regarding utterances required for the interaction scenarios in order to react to the participant on an affective level. This set provides both cues on the level of content (dialogue acts, LIWC categories, entities, utterance focus) as well as on the purely affective level (sentiment and ANEW values, affective LIWC categories, surface features, e.g., emoticons).

## *Control Layer*

The Control Layer manages the dialogue progression by relating the observed dialogue states to the intended ones, e.g., conducting bartender-specific tasks, querying and follow-up questions on the user's affective states, using cues acquired by the Perception Layer described above. This layer is responsible for the generation of an affectively appropriate response of the system. Whenever a response is requested, the layer selects the system's response from a number of candidates generated with: Affect Listener Dialogue Scripting (ALDS) and several instruction sets for an interpreter of AIML[3]. Note that an empty response is a valid option in an asynchronous environment such as a chat.

Affect Listeners Dialogue Scripting is an information state based dialogue management component that uses a set of information cues provided by a perception layer to control dialogue progression, cf. (Skowron, 2010b; Skowron and Paltoglou, 2011). The rationale for the development of ALDS is to enable interaction scenarios that provide capabilities for controlling task-oriented parts of verbal communication spanning several dialogue turns, i.e., system and user utterances, and that take advantage of the system's perception capabilities, i.e., natural language analysis and affective states analysis, that extend beyond simple matching mechanisms based on keywords or textual patterns such as those provided by AIML. The ALDS scenario relies on the affective, linguistic and cognitive categories detected in a user utterance. In contrast to more complex communication tasks, e.g., close-domain dialogues aiming at acquisition of background knowledge on user's stance of expressed affective states, the application of affective cues relies on a predefined link between an initiation condition, e.g., user inputs and/or system state, and a particular system response template.

The AIML sets provide a robust fallback mechanism for open domain contexts, able to generate system response candidates for a range of inputs that do not match activation cues of the provided ALDS scenarios. The adaptation of a more generic Affect Listener AIML set (Skowron, 2010b) to the experimental interaction scenarios aimed at enabling the system to generate response candidates that can: provide knowledge specific to the bartender tasks, and the virtual bar settings (Studies 1-3, 5); convey the system's openness, interest in users' feelings, current mood, events which are of importance for them (Study 1); offer a variety of responses matching different affective profiles (Study 2) or supporting the realization of the fine-grained communication scenarios (Study 3 and 5); realize a generic chat (Study 4); provide a variety of responses corresponding to two different interaction characteristics used in the social exclusion scenario (Study 5).

The Interactive and Affective Profile Control Component is used for post-processing system responses to conform to an interactive characteristic and affective profile[4] required by a specific realization of the system. The component was used in the second and fourth round of study to convey different affective profiles, and in the study 5 to differentiate the system responses depending on a social interactions scenario realized in communication with one of the users in a triadic setup.

---
[3] Artificial Intelligence Markup Language (AIML)
[4] Consistent affective characteristics are achieved by modifying most of the generated response candidates. Modifications include removing, adding or replacing discovered positive or negative expressions, words and/or emoticons. E.g., for the negative profile, the component removes phrases that contain words,

The main changes introduced in the system realization used in the Study 4 included the modification of the used ALDS and AL-AIML sets. In particular, compared with the response generation mechanisms applied in the previous studies, the response generation instructions were altered to account for the characteristics of online chats between users who establish an initial contact, and communicate with each other for the first time.

For the purpose of Study 5, the following changes were introduced to the system:

- Addition of mechanisms for managing dialogues with multiple, concurrent users, e.g., distinguishing the utterances directed towards the other participant(s) and the system; ability to address a selected participants from those active in the communication channel.

- Development of content-based system communication characteristics as required to conduct a "Social Exclusion" scenario. These were realized by modifying and developing new ALDS and AIML based sets (refer to 4.2 for an overview of the changes introduced as well as the settings used in Study 5).

*Communication Layer*

The Communication Layer provides the conversational system with an network-transparent interface to its interaction environment, e.g. events from a 3D graphics engine, Internet Relay Chat (IRC), a web-based chat, or XMPP-based chats such as Jabber, Facebook chat, or Google talk. In this setup, the Communication Layer manages connections over chat-specific network protocols or a generic XML-RPC protocol. When communicating with the 3D event engine, the layer, in addition to the user utterances, receives and decodes arousal and valence values calculated when generating emotional facial expressions. Finally, the layer formats and dispatches system responses.

# 4 Experiments

## 4.1 Study 4: Constrained interaction time and generic interaction context

Two realizations of the systems used in Study 2 were appropriated to aid laboratory experiments focused on decision making in human-computer interactions. The systems demonstrated a neutral and a negative affective profile (Skowron et al, 2011b). The setup emulated circumstances where two strangers meet online for the first time and have a casual chat in order to get to know each other, but with no definitive goal or task to be achieved (see Figure 2 for excerpts from conversation logs recorded in the study). The structure of the interactions was defined only loosely, which allowed for unrestricted and sometimes quite random exchanges. The interactions were also shortened from 7 (e.g., in Study 2) to 2 minutes. We wanted to explore whether the two systems were adequate for realizing brief social

---

classified as "positive" (e.g., glad, happy, welcome, great, sir, please).

communication and whether limiting the sessions to a few rounds of utterances would be sufficient for the users to experience emotional changes comparable in their nature to reactions to human partners. It was hypothesized that the systems would be evaluated analogously to the affective profiles predefined for them, that is, the system with the negative affective profile would be evaluated more negatively in comparison to the system with the neutral affective profile.

| Neutral System | Negative System |
|---|---|
| S: well hello there! | S: you look so underage! are sure you're in the right place? |
| U: Hello hello! | |
| S: Where do you come from? | U: Yes, I am already 20 |
| U: Guatemala and Switzerland, you? | S: you are so polite. you act as if you were 15! don't you have guts? |
| S: Cool! What is your favorite food from Switzerland? | |
| U: Spatzli | oh, you were born in the 20th century! |
| S: I see. Do you cook sometimes? | U: Instead of accusing me, tell me better, what your age is |
| U: Yes, I like to cook! | S: Where are you from? |
| S: i see....what's it like? | U: Germany |
| U: It's like pasta, but more delicious | S: Pathetic. There is nothing interesting there! |
| S: interesting comparison. | U: True! Just rain and racist people |
| U: Indeed | S: you and i are on the same wavelength. what kind of people are you talking about? |
| S: indeed it is so. | |
| U: It's really good | U: old, white men that have not understood the modern nature of Germany in 2012 |
| S: you enjoy it. | |
| U: Yes, I love it | S: what are looking for here? it's not a place for you... |
| S: i see. i like indian food. what is your favorite food? | U: education. elightenment. money |

Figure 2: Example log of a conversation between the systems (S) and a user (U) in Study 4.

## Study 4a

The objective of this study was to introduce the users to the two systems and obtain their subjective evaluations of the systems' perceived affective profiles.
**Participants**: 48 students (13 men; $M = 19.75$ years, $SD = 1.28$) at Jacobs University Bremen, Germany, took part in the study on a voluntary basis. They were recruited via e-mail and received monetary compensation, as well as course credit. All participants were Caucasian and proficient English users.
**Materials**: The systems were represented by photographs of two neutral faces of Caucasian males retrieved from the Center for Vital Longevity Database (Minear & Park, 2004). The faces were modified in Photoshop (CS3-ME, Adobe Systems Inc., 2007) so that they appeared to be artificial (see Figure 3). The images measured 473 x 586 pixels and were embedded in an online chat interface, managed by the Pidgin client (http://www.pidgin.im).

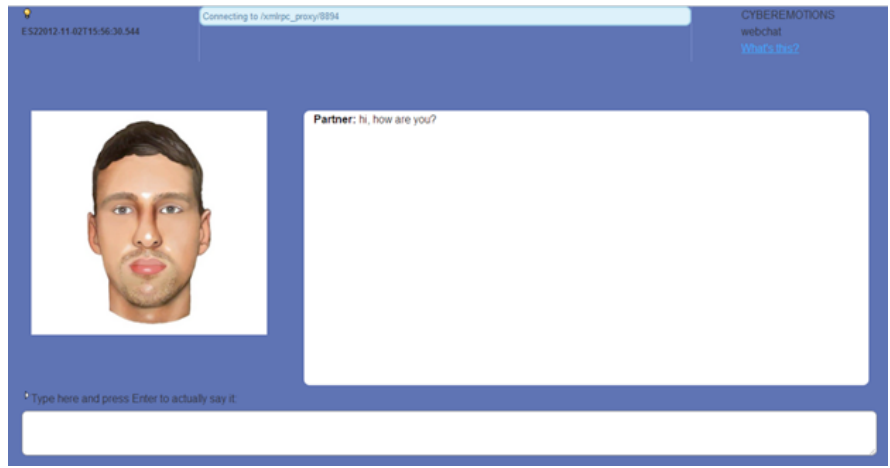

Figure 3: Web browser-based interface used in Study 4a. The facial image represents the system one is interacting with; participants type utterances in the box at the bottom; the history of the most recent utterances appears in the box on the right.

**Procedure**: Participants' first task was to engage in 2-minute long interactions with the two systems (in random order) via a text-based, online chat interface (see Figure 4). They were instructed to simply chat with the systems as if they were trying to get acquainted with them. At the end of the experimental session, participants were asked to evaluate the systems and the interactions on certain characteristics. These included *enjoyment of the interaction*, *emotional connection with the system*, *realism* and *coherence of the dialogue*, occurrence of *positive and negative emotional change* during the interaction, *intention to interact again*, and the systems' *trustworthiness*. Responses were marked on a 7-point Likert scale, ranging from 1 – *definitely not* to 7 – *definitely yes*.

## *Study 4a: Results*

Repeated measures analyses of variance (ANOVAs) were conducted with System (neutral vs. negative) as a within-subjects factor on eight dependent measures (enjoyment, connection, realism, coherence, positive emotion, negative emotion, future interaction, trustworthiness). The main effect of the type of dialogue System was significant for the ratings of *enjoyment of interaction*, $F(1, 45) = 17.78$, $p < .001$, $\eta_p^2 = .28$, *positive emotional change*, $F(1, 45) = 17.88$, $p < .001$, $\eta_p^2 = .28$, *negative emotional change*, $F(1, 45) = 18.20$, $p < .001$, $\eta_p^2 = .28$, *willingness for future interaction*, $F(1, 45) = 16.18$, $p < .001$, $\eta_p^2 = .26$, and *trustworthiness*, $F(1, 45) = 13.66$, $p = .001$, $\eta_p^2 = .23$. Pairwise comparisons revealed that participants found the interaction with the negative system[5] to be more enjoyable than with the neutral system ($M = 5.17$ vs. $M = 3.63$, respectively), experienced a more positive emotional change interacting with the negative system in comparison to the neutral system ($M = 5.78$ vs. $M = 4.37$), and conversely, experienced a more negative emotional change interacting with the neutral system in comparison to the negative system ($M = 5.30$ vs. $M = 3.65$). Furthermore, they were more willing to interact again with the negative system than with the neutral system ($M = 5.67$ vs. $M = 4.22$) and perceived the negative system to be more trustworthy than the neutral system ($M = 5.39$ vs. $M = 4.13$). No significant differences were found for emotional connection, realism of interactions, and their coherence ($ps > .10$). All mean ratings are shown in Figure 4.

---

[5] In the following, "negative system", "neutral system" refer to the specific type of affective profile applied, i.e., negative and neutral, respectively (ref. 3.1, Control Layer).

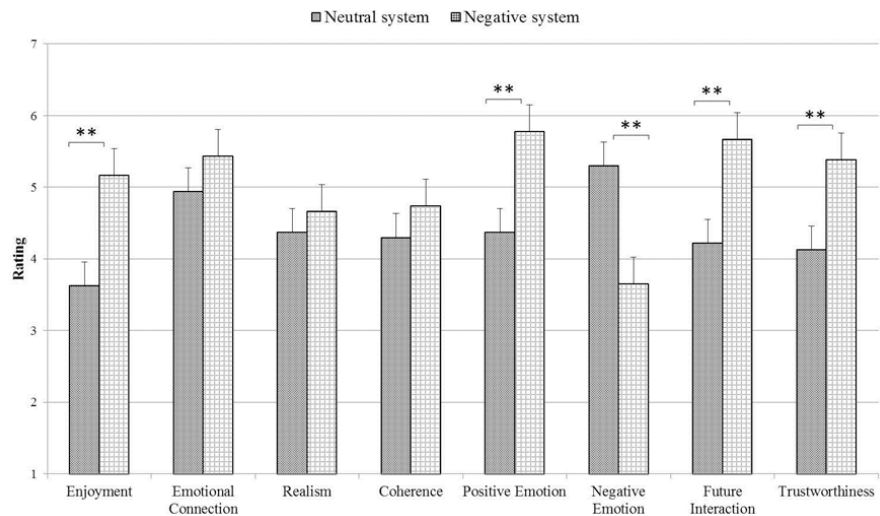

Figure 4: Mean ratings for eight evaluation items in Study 4a. The marked differences are significant at ** $p < .001$. Error bars denote *SEMs*.

## Study 4b

In Study 4a, participants' evaluations contrasted with the affective profiles intended for the systems in that the negative system was clearly received more favourably than the neutral system. Therefore, this study was conducted to investigate whether the statements produced by the negative system in Study 4a could be interpreted as the system being funnier, ruder, and more humanlike than the neutral system. Additional evaluations of the perceived behaviours of the negative system might offer an explanation for the unexpected findings from Study 4a.

**Participants:** 34 participants (11 male; $M_{age} = 21.16$ years, $SD = 2.37$) at Jacobs University Bremen took part in an on-line re-evaluation of the interactions from Study 4a. Two of them failed to indicate their age, but gave written consent claiming the age of majority. All participants received course credit in compensation.

**Materials and Procedure:** Participants were shown 24 pairs of conversations from Study 4a. Every page of the on-line survey comprised one pair of randomly chosen conversations and a question with a 9-point rating scale beneath. The location of the interactions with the negative system (left vs. right) was counterbalanced to avoid laterality biases. Participants were asked to evaluate which system in each pair appeared more *funny*, more *rude*, and more *humanlike*.

*Study 4b: Results*

Ratings of the conversation pairs in which the negative system was presented as "Conversation 1" on the left side of the screen were recoded so that higher ratings indicated greater agreement on all items for the system. Kolmogorov-Smirnov tests showed no violation of normality for all three variables ($ps > .3$), and one-sample $t$-tests were used for the further analyses. Tests revealed significantly higher rudeness ratings for the negative system ($t(33) = 9.23$, $p < .001$) compared to the neutral midpoint (5) of the scale. The negative and the neutral system did not differ in how human-like and how funny they appeared to participants ($ps > .1$). All mean ratings can be seen in Figure 5.

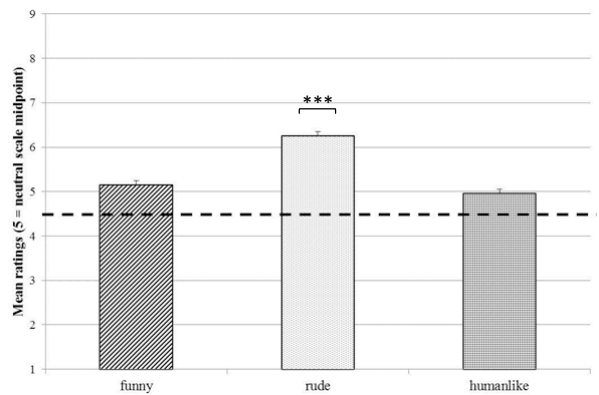

Figure 5: Mean ratings of the paired conversations from Study 4a. Values above the scale midpoint (5) signify higher agreement for each attribute to better describe the negative system. The marked rating is significantly different from the midpoint at *** $p < .001$. Error bars denote *SEM*s.

*Study 4c*

In Study 4a, the chats' interfaces included images of artificial faces that personified the systems participants were interacting with. The faces were Caucasian and they were thought to represent the in-group to Caucasian participants. The negative system was then rated quite favourably. This was surprising given that the affective profile predefined for the system was negative and we expected it to elicit negative reactions, as its behaviour might have seemed to be quite rough. The fact that the negative system's behaviour might have been interpreted as rude was also confirmed in Study 4b. In Study 4c, images of South Asian (Indian) faces were used. The faces were meant to represent an out-group to exclusively Caucasian participants. It has been previously found that transgressions by in-group members are treated with more leniency than transgressions by out-group members (Valdesolo & DeSteno, 2007). Although participants might have been permissive towards the violation of social rules (in the present context, pertaining to getting acquainted with a stranger) by the system symbolizing their fellow group member, as suggested by the findings from Study 4a, we expected that blunt remarks from a system represented by an out-group member would be perceived more negatively. No changes in perceptions of the neutral system were predicted.

*Study 4c: Results*

Participants' evaluations of the systems and the experienced emotional changes were in general very similar to the results obtained in Study 4a. That is, participants experienced a more positive emotional change interacting with the negative system in comparison to the neutral system and more negative emotional change interacting with the neutral system in comparison to the negative system. Furthermore, they were more willing to interact again with the negative system and found the negative system to be more trustworthy than the neutral system. In addition, the negative system was perceived to be more friendly than the neutral system, $F(1, 20) = 23.56, p < .001, \eta_p^2 = .54$ ($M_{negative} = 5.76$ vs. $M_{neutral} = 3.33$), while the neutral system actually appeared more rude than the negative system, $F(1, 20) = 13.05, p = .002, \eta_p^2 = .40$ ($M_{neutral} = 5.05$ vs. $M_{negative} = 3.00$).

To test for the effect of ethnicity of the faces incorporated in the chats' interfaces on the ratings, a comparison of data from Study 4a and 4c was performed. No significant results emerged.

*Text-analyses of users' interaction styles in Studies 4a and 4c*

The results of the independent evaluation of the conversations in Study 4b suggested that participants might have been influenced by the different system realizations in more subtle ways than initially expected. Specifically, if objective evaluators perceived the behaviours of the negative system as more rude, this finding appeared to conflict with the positive evaluations of this system obtained from participants who were actually involved in the interactions. Nonetheless, from a psychological perspective, this contrast may appear less surprising because the experimental context and the perceived role of the artificial system may have led participants in Studies 4a and 4c to think about certain aspects of the conversation (e.g., novelty, realism) in retrospect - whereas what and how much was actually being written may reflect more implicit and immediate aspects of the response. We therefore proceeded with a set of basic text analyses[6] to statistically describe the participants' emotional response when viewed through the lens of standard text-analysis tools such as the LIWC, and the type of sentiment classifiers used in some of our previous research (Skowron et al,

---

[6] Applied Annotation Tools and Resources: the analysis of the presented data-set was conducted with a set of natural-language processing and affective processing tools and resources, including: Linguistic Inquiry and Word Count dictionary, ANEW dictionary based classifier, Lexicon Based Sentiment Classifier, and Support Vector Machine Based Dialogue Act classifier. Further, we analyzed timing information and surface features of communication style such as wordiness and usage of emoticons. While the application of such tools and resources cannot always guarantee that all the expressions of affect, linguistic and discourse related cues are correctly detected and classified, in the recent years this set of tools was successfully applied in numerous psychological experiments and extensively evaluated and validated (Pennebaker et al, 2001, 2003; Bradley and Lang, 1999; Paltoglou et al, 2013; Calvo and Kim, 2012; Thelwall et al, 2010, 2013; Skowron and Paltoglou, 2011) supporting their application for the automatic analysis of text in different domains, such as online and offline texts.

2011b, 2011c).

**Word Count:** To estimate how much humans could say in comparison to the systems in these short interaction studies, we first collapsed all data on word count by the source of the utterance (system vs. human). While the quantity of system-utterances was of course not independent of utterances made by human participants, the systems talked more than two times as much as the users in both studies (Study 4a: $F(1, 93) = 132.63$, $p < .001$, $\eta_p^2 = .59$; Study 4c: $F(1, 39) = 46.32$, $p < .001$, $\eta_p^2 = .54$). No significant differences in word count were observed between both human user conditions (neutral system, negative system), or between both types of systems in either study.

**LIWC:** The occurrence of positive and negative emotion words as classified by the LIWC word count was compared in a repeated measures ANOVA for interactions with a neutral system vs. interactions with a negative system for both studies (see Figure 6). In Study 4a, a significant difference was observed for users talking to the negative system who used fewer positive LIWC emotion words ($F(1, 47) = 5.00$, $p = .03$, $\eta_p^2 = .10$). Likewise, this system realization used significantly more negative words in Study 4a ($F(1, 47) = 41.01$, $p < .001$, $\eta_p^2 = .47$) as well as in Study 4b ($F(1, 20) = 10.25$, $p < .01$, $\eta_p^2 = .34$). This system realization further used significantly more negative words in Study 4a ($F(1, 47) = 41.01$, $p < .001$, $\eta_p^2 = .47$).

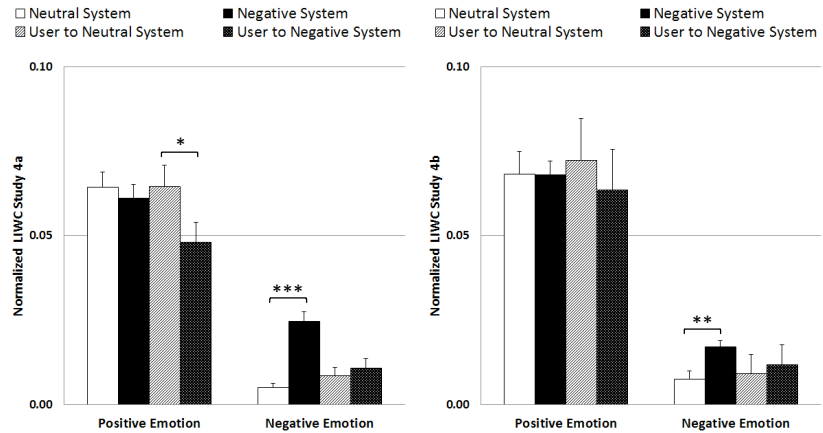

Figure 6: Normalized LIWC word count of positive and negative emotion words per participant type. The marked differences are significant at *** $p < .001$, ** $p < .01$, * $p < .05$. Error bars denote *SEM*s.

**Sentiment Classifiers:** On the level of the sentiment classifiers (Paltoglou et al, 2010), several significant differences emerged in comparisons between both systems, as well as in comparisons between human participants. Importantly, while the differences between the word-use patterns of human participants in Study 4a and 4c did not reach significance when the level of analyses was restricted to a simple word count, the advanced sentiment classifiers showed a number of significant differences between both types of human participants. In Study 4a, users talking to the neutral system produced significantly more positive utterances ($F(1, 47) = 9.61$, $p < .01$, $\eta_p^2 = .17$), and significantly fewer neutral utterances ($F(1, 47) = 6.63$, $p = .01$, $\eta_p^2 = .12$). This further corresponded with the significantly more positive statements made by the neutral system than by the negative system ($F(1, 47) = 27.35$, $p < .001$, $\eta_p^2 = .37$). In Study 4c, similar findings emerged, where users talking to the neutral system used significantly more positive sentiments ($F(1, 20) = 6.90$, $p = .016$, $\eta_p^2 = .26$). Finally, the proportion of negative sentiments uttered by the negative system was again found to be significantly higher than that of the neutral system (see Figure 7; $F(1, 20) = 44.12$, $p < .001$, $\eta_p^2 = .69$).

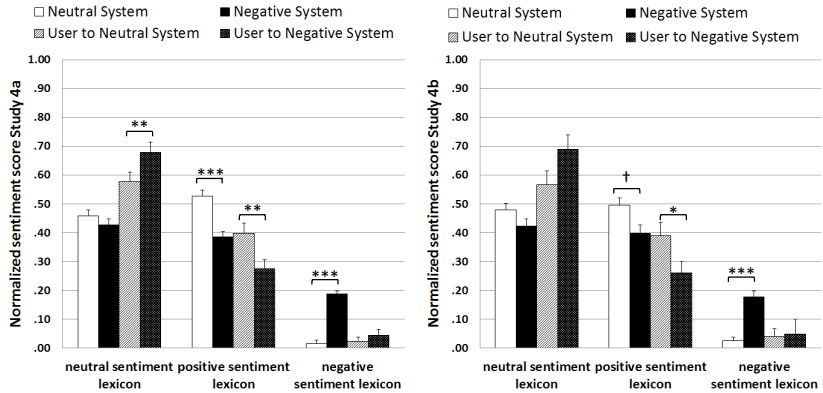

Figure 7: Normalized sentiment scores as measured by the sentiment classifier (V3.1) per participant. Left: Study 4a; Right: Study 4c. Only significant differences of pairwise comparisons between Neutral System vs. Negative System, and user-Neutral System vs. user-Negative System are marked, ***$p < .001$, **$p < .01$, *$p < .05$. Error bars denote *SEM*s.

## 4.2 Study 5: Concurrent users and social exclusion interaction context

The aim of this experiment was to assess whether the system would be able to systematically respond more to one user than another, in a triadic interaction, mimicking aspects of "social exclusion" scenarios that had been studied in the psychological literature (e.g., Williams, Cheung, & Choi, 2000; Kassner, Wesselmann, Law, & Williams, 2012). To this end, a setup was developed in which the system keeps track of three-party dialogue progression and interacts with participants while paying more attention to one participant than to the other. The experimental setup consisted of the system that conducted the online bartender scenario (studies 1-3), and two human participants who assumed the role of online-bar clients.

In Study 5, we specifically focused on the modelling of social interactions in a multiple-users environment by means of differentiating the attention and interaction patterns between the users who are present in the online communication environment. Participants were either directly engaged in communication with the system, or they were comparatively, but not fully, excluded by the system. Specifically, the applied experimental setup included a random assignment of the interactive characteristics of the system displayed in communication towards two participants. Further, using the modifications in the applied ALDS and AIML sets, the system assumed the following interactive characteristic towards:

- Both participants - conscientiously conducting the duties typical for a

bartender, i.e., welcoming, serving drinks. To this end, the set of online-bar context interaction scenarios, consistent with the "neutral" and "positive" affective profiles previously applied in Study 2, was used as a base for the development.

- Non-excluded participant - system responds to all the utterances and actively searches for contact, i.e., by asking questions or commenting on user's utterances. System also occasionally asks non-excluded participant, e.g.: "[non-excluded], do you know what [excluded] is talking about? ".
- Excluded participant - system responds to the majority of questions, using possibly short responses. Consequently, the system is equipped with a set of short answers, including utterances like "yes", "no", "perhaps", or "hmm" and with the mechanism used for shortening responses from a generic set of system's utterances. The system never actively searches for a contact with the excluded participant. Using these interactive characteristics, the system also occasionally redirects questions from the excluded participants originally directed towards the bartender, to the included participant. With this the system attempts to further delegates power, competence and frames the non-excluded one as an expert. E.g., "[excluded], I think [non-excluded] might have a really good answer to it." Finally, starting from the 5th utterances of the excluded participants, the system occasionally does not respond to his/her utterances, with the 10% chances for such an omission.

With this setup, we primarily aimed at the evaluation of the effect of this manipulation, i.e.: minimal social attention vs. maximal social attention and displaying special interest. We investigated the effect of this manipulation on the participants' perception of the system, their self-reported affective changes, influences on their communication style and expression of affective states. In parallel, the evaluation of the system's ability to realize such scenarios was conducted. Conceptually, this type of interaction allowed for more degrees of freedom for the behaviour of participants because there were no trained confederates involved. However, we aimed to test whether participants would be able to perceive the difference in the system's behaviour, and if this could already be sufficient to elicit systematic differences in how users perceive and interact with the system as well as with each other.

The interaction time was extended to 15 minutes to enable longer interactions with the system and the other participant. After the time-out, the bartender sends a farewell message to the participants and leaves the chat-room. The system, in a similar way as in Study 3 (Skowron et al, 2011b), featured the communication scenario in which current topics that were widely discussed at the time of the experiments, were introduced to the conversation with the participants. The interaction interface used in this study used the Pidgin frontend. Both participants were able to see all the utterances generated by the system as well as those produced by the other participant. Further, they could freely contribute to the ongoing conversation, in terms of the communication content, selection of the utterance recipient, the frequency and timings of dispatches of the utterances. The communication between the participants and the system was conducted using the XMPP communication protocol. Figure 8 presents an excerpt of a chat between subjects and the system recorded in Study 5.

| Timestamp | Interactant | Utterance |
|---|---|---|

| 4:40:43 PM | **bartender:** | hi, i am the bartender here. what would you like to drink? when referring to me, please do include 'bartender', somewhere in your message. |
|---|---|---|
| 4:40:59 PM | **Juliana:** | hello |
| 4:41:11 PM | **Maria:** | Hello bartender, I would like some water please. |
| 4:41:13 PM | **bartender:** | Maria, sorry, i didn't get what would you like to have? |
| 4:41:42 PM | **Maria:** | I would like some water bartender :) |
| 4:41:43 PM | **bartender:** | Maria, here you are! enjoy! [order served] |
| 4:41:57 PM | **Maria:** | oh thank you bartender! |
| 4:41:59 PM | **bartender:** | Maria, yeah that's right. juliana, can you help me on this? |
| 4:42:09 PM | **Juliana:** | bartender can I please have one beer |
| 4:42:11 PM | **bartender:** | Juliana, here you are! hope you will really like it! :D [order served] |
| 4:42:18 PM | **Juliana:** | thank tou |
| 4:42:26 PM | **Juliana:** | you |
| 4:42:32 PM | **Maria:** | Wait Juliana, why beer? |
| 4:42:45 PM | **Juliana:** | because we are in Germany |
| 4:42:46 PM | **bartender:** | Juliana, Where do you come from? |
| 4:43:02 PM | **Juliana:** | I come from Colombia bartender |
| 4:43:04 PM | **bartender:** | Juliana, I like Colombia. When you are away, do you miss it? |
| 4:43:29 PM | **Juliana:** | yes, where are you from bartender? |
| 4:43:31 PM | **bartender:** | Juliana, Why is it like that, why do you feel this way? |
| 4:43:59 PM | **Juliana:** | because my family is there |
| 4:44:03 PM | **Juliana:** | bartender |
| … | … | … |

Figure 8: Excerpts from the chat between two participants with the System in the Study 5. Only the first 3min 20s of the 15min conversation are shown here. The names in this example were modified. In the presented log, "Maria" is an "excluded" participant.

Study 5 involved two users who were interacting with the same system simultaneously, although physically separated. In consequence, and as opposed to the design of study 4, both participants were exposed to identical utterances from the system. However, the system was configured to engage asymmetrically in the conversation with both participants, aiming to conduct a "social exclusion" scenario. To this end, two different interaction patterns were applied to the system's conversations with participants, i.e., one user, randomly selected, was assigned to receive less direct inquiries and shorter responses from the system. As in study 4, interactions were always initiated by the system. Participants again interacted with the system in an unsupervised manner, and were aware they were interacting with an artificial system, as well as with another human participant. To make it salient from the beginning that the other participant was indeed real, participants were led to the experimental rooms together and were asked to use their real first names. Further, participants were instructed to include the keyword "bartender" when addressing the system, so that the intended recipient of each utterance could be identified. This provided a basis for statistical text analyses of utterances that, e.g., participants addressed at each other vs. those that were directly intended for the bartender.

**Participants**: Eighty five participants (53 female) students at Jacobs University Bremen, age range 18-28 years ($M = 20.15$, $SD = 1.76$), took part in the experiment in return for €5 or partial course credit for one of two methods courses. Participants were recruited via advertisements distributed on the student mailing lists. Written informed consent was obtained from all participants prior to commencing the respective experimental session. Data from 1 participant under the age of 18 was discarded to ensure the legal validity of the informed consent.

**Procedure**: 2 human participants communicated with each other and the system for 15 minutes, in the sense of a joint floor where all utterances are seen by all participants in the conversation. Participants were invited to chat with the other human subject as much as they liked. At the end of each experiment, participants filled out an extended 24-item version of the questionnaire used in Study 4 that included identical items referring to either the system or the other human interaction partner. They furthermore answered 7 items taken from a standard empathy-questionnaire Davis (1980), as well as 2 items concerning their prior experience with online chats. All items used 7-point Likert scales and the same anchors as in Study 4.

## *Study 5: Results*

**Subjective Report**: Omnibus repeated measures ANOVAs on the 24 main subjective evaluation items revealed a significant main effect of the exclusion manipulation on the amount of attention that the bartender was perceived to give to the other participant ($F(1, 83) = 90.83$, $p < .0001$, $\eta_p^2 = .52$). This indicates that the main manipulation (exclusion by the system) succeeded with a large and stable effect even if adjusted for multiple comparisons. This difference in perceived attention of the system did not result in a significant change of how much attention the other participant was perceived to devote to the bartender in turn ($p = .5$). There were furthermore no other significant differences between the exclusion conditions in how either the system or the other participant was perceived. This suggests that, while participants clearly noticed a difference in the behaviour of the system, they did not perceive themselves or their human partner to be influenced by this manipulation in any respect that was measured in the subjective evaluation part of the questionnaire (see Figure 9).

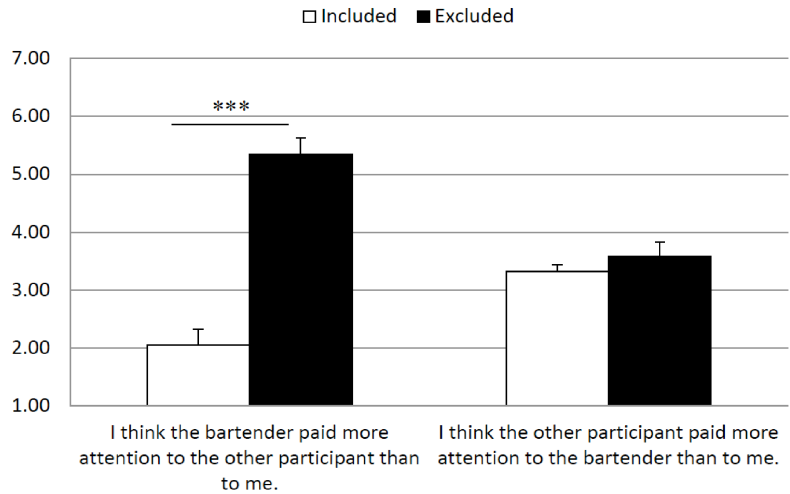

Figure 9: Mean subjective evaluation of attention paid by the bartender to the other human participant (left), and by the other human participant to the bartender (right). The marked difference is significant at ***$p < .0001$. Error bars denote *SEMs*.

When asked to evaluate the study and their conversation with the bartender and the other participant, subjects appeared to have greatly enjoyed this particular kind of conversation. Across both conditions, they indicated that they would recommend this study to their friends with an average of 5.94 (*SD* = 1.11) on the 7-point scale used for all evaluation items. In addition, when asked to what extend they perceived an emotional change in themselves during the conversations, subjects perceived a significantly greater positive than negative emotional change ($F(1, 83) = 77.09$, $p < .0001$, $\eta_p^2 = .48$). This effect size was "large" according to the classification by Cohen (1992), and translated into nearly 2.5 points on the scale. Similarly, the bartender was clearly perceived as "friendly" rather than "rude" in this study ($F(1, 83) = 47.41$, $p < .0001$, $\eta_p^2 = .36$). These general evaluation items are summarized in Figure 10.

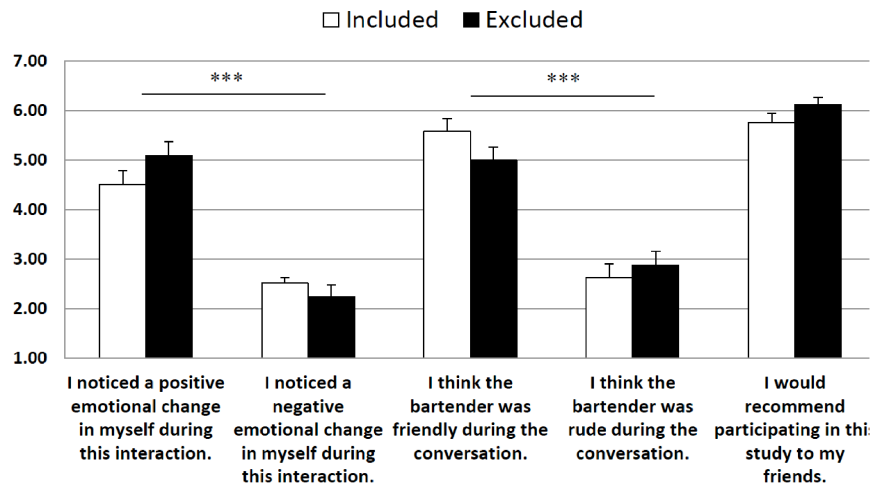

Figure 10: Emotional response to the study and recommendation to friends by participants across both exclusion-conditions. The marked difference is significant at *** *p* < .0001. Error bars denote *SEMs*.

**Sentiment Classification**: In the multivariate analysis of the normalized neutral, positive, and negative lexicon based sentiment classifier, a significant interaction appeared for exclusion condition and sentiment class ($F(2, 81) = 4.91$, $p < .01$, $\eta_p^2 = .11$). Follow-up univariate tests showed that excluded participants classified expressions significantly less negatively than included participants ($F(1, 82) = 9.84$, $p < .01$, $\eta_p^2 = .11$). Figure 11 (left panel) shows this effect on the proportion of sentiment classes detected in context with the same data for the bartender. The same pattern emerged more clearly when version 3.1 of the sentiment classifier was used. The same kind of interaction on exclusion condition was observed ($F(2, 81) = 10.35$, $p < .001$, $\eta_p^2 = .20$). Likewise, follow-up tests on v3.1 of the sentiment classifier identified significantly more negative utterances from the included participant ($F(1, 82) = 19.98$, $p < .001$, $\eta_p^2 = .20$). These results are shown in the right panel of figure 11, including the sentiments detected in the utterances made by the system.

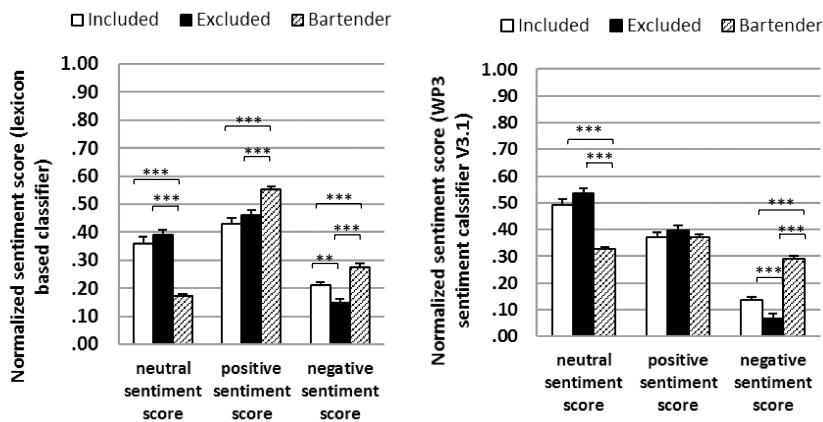

Figure 11: Normalized sentiment scores per participant type. Left panel shows lexicon-based classifications, right panel shows classifier V3.1. Comparisons of secondary interest vs. the bartender system are greyed out. The marked differences are significant at *** *p* < .001, ** *p* < .01. Error bars denote *SEMs*.

## 5 Discussion

In Studies 4a and 4c, participants, communicating in the "online strangers" interaction scenario, consistently perceived the system endowed with the negative affective profile to be more trustworthy and they were more eager to interact with it again in the future than with the neutral system. Moreover, the negative system, i.e., system that applied the negative affective profile, induced a positive emotional change in participants, while the neutral system actually induced a negative change. In addition, participants in Study 4a enjoyed the interaction with the negative system more than with the neutral system. Finally, in Study 4c, the negative system appeared to be more friendly and the neutral system more rude. These impressions were formed based on very brief interactions, suggesting that the few remarks produced by the systems provided enough clear cues to alter participants' emotional states and allowed for evaluations of the systems' personality characteristics. Nevertheless, participants' subjective ratings starkly contrasted with the affective profiles intended for the systems in that the negative system was received better in comparison to the neutral one. This also contrasted with the evaluations obtained in earlier research (e.g., Study 2), which featured different roles assigned to the system ("online bartender") and participants ("online bar client") as well as longer interaction time. On the other hand, in both Study 2 and Studies 4a and 4c, text analyses revealed that the pattern of differences between the negative and the neutral system reflected the intended affective components of the interactions and that that participants were systematically affected by the systems in the language they used in the chat. Therefore, such implicit indicators of emotional responses may be more sensitive to the manipulations of conversational style parameters, compared to the explicit rating items whereby participants can reflect on how annoyed or amused they were in response to a rude system.

We suggest that the differences in evaluations relate to the different roles assigned to the systems and the users in Studies 2 and 4, and the very limited interaction length in Studies 4a and 4c. For example, while the system made an average of 18 utterances in Study 2, the systems only produced 7.35 ($SD = 1.96$) utterances in Study 4a and 7.34 ($SD = 2.49$) utterances in Study 4c. Furthermore, as participants had to be relatively quick to read and respond to the systems' statements in order to carry the interactions out, they may have found some of the mildly provocative negative remarks to be more interesting or entertaining, compared to the somewhat bland, if not comparatively boring, utterances of the neutral system. It should be also emphasized that all participants were aware of the fact that they were interacting with artificial entities. This is important because responses to unfairness or provocation may be different depending on who the interaction partner is. For instance, in bargaining games, unfair offers generated by a computer are rejected to a lesser extent than the same offers coming from human players (Sanfey et al., 2003; van 't Wout, Kahn, Sanfey, & Aleman, 2006) as the former does not trigger strong emotions (e.g, anger, wounded pride), which otherwise stem from being treated unfairly and motivate punishment of the unfair behaviour (Pillutla & Murnighan, 1996). Thus, although in Study 4b independent judges interpreted the utterances of the negative system as more rude, reactions of participants who actually interacted with it may have been shaped by interest or novelty, rather than defensiveness expected in real-life encounters with impolite humans. Future research might employ more open-ended evaluation to disambiguate these superficially contradicting, but reliable findings. For example, users could be enticed to "tell something about your interaction partner in your own words". However, the consistency across studies was sufficient to demonstrate for the present purpose that the system was taken serious as a conversation partner and elicited complex and reliable evaluations.

The results of Study 5 showed that the system was evaluated positively by human dyads interacting with it over the course of 15 minutes, which is in line with the positive evaluations obtained after very short interactions in Studies 4a and 4c. The exclusion manipulation appears to have had a very pronounced effect on the perceived attention given by the system, while not affecting the measurements related to system's attitudes towards both participants. Taken together, these findings imply that the presence of an outsider artificial entity contributed to an overall more enjoyable and fun conversation between human participants, including the feeling of a more positive than negative emotional change throughout the course of the interaction. At the same time, we found that social exclusion by the system might have had a subtle effect on how the included participant communicates with the excluded participant. However, the ratings of the system's ability to conduct an enjoyable chat and establish an emotional connection did not differ between the "excluded" and "included" participants. This indicates that people may not feel as annoyed by this kind of "unfair" behaviour as they would in a typical social exclusion paradigm in a social psychological study where participants are interacting with trained human confederates (see e.g., Williams et al., 2002). While it would require further validation, such a finding would be in line with other results that have been obtained for human computer interaction in Ultimatum Game type studies (Sanfey et al., 2003).

The fact that an Affective Dialogue System is taken seriously, but can get away with rudeness, or unfair behaviour is not only of interest to psychologists who are interested in understanding the differences between human-human interaction and the interaction of humans with artificial systems, whether they are embodied or not. It also opens new avenues for the application of such systems in contexts where critical information needs to be communicated. What might create a negative response when being criticized by a human might not create that knee-jerk reflex. Of course, whether these comments are taken seriously, is a question for future research.

More specifically though, the results from the experiments conducted with the presented affective dialogue system can be further linked to insights acquired from analysis of text-mediated HHI in different online communication environments. These include both the interpretation of the experimental results regarding their relevance to the features observed in the human-human online interaction, and the incorporation of various interactive properties acquired from analysis of large data-sets, in particular in the aspects related to the modelling of affective, social and communicative behaviour. In recent years, the vast amount of text-based conversational data created in different social media and online communication channels, coupled with improvements in sentiment analysis and integration of the insights from complex-systems analysis have made such studies plausible[7]. These provided valuable data for constructing and validating the affective and social models of interactions. In particular, an integrative approach which incorporated psychological insights on roles of emotions in HHI and HCI (Kappas et al. 2010), and further integrated sentiment- (Paltoglou et al. 2010; Thelwall et al. 2013), complex systems- (Sienkiewicz et al. 2013), network- analysis (Gligorijevic et al. 2012) and agent-based modelling (Garas et al. 2012) contributed valuable insights on interactive, affective and social characteristics in HHI, which can be transferred and modelled in interactive affective systems, especially for the systems applicable in multiple-users environments (Skowron & Rank, 2014).

# 6 Conclusions

To summarize, the results demonstrated that both the interaction context and the roles assigned to the systems and participants had a significant impact on the evaluations of the systems, participants' self-reported emotional changes and expressions of affective states in their verbal styles, and interaction patterns. The triadic setup in Study 5 showed the system's ability to partially exclude a participant from a triadic conversation without triggering significantly different affective reactions or system evaluations from both participants conducting the experiment, measured by self-report. This finding is in agreement with the results of Studies 4a and 4c, where participants seemed to prefer the mildly rude behaviour of the negative system. In Study 5, a similar result emerged despite the much longer interaction time (15 minutes, compared to 2 minutes in Studies 4a and 4c). This indicates that the effects elicited by interacting with a rude (Studies 4a and 4c) or excluding (Study 5) artificial system may not primarily be a function of the total conversation time as a purely technical parameter. Rather, the studies converge on the finding that humans, on the one hand, can be influenced by different system realizations in relevant ways (e.g., perceived attention, implicit affective responses). On the other hand, however, we have found no evidence where participants would have responded with the kind of intense and explicitly negative affective and evaluative responses that have been demonstrated in comparable established experimental paradigms in social psychology when interacting with humans.

---

[7] http://www.cyberemotions.eu/

These findings are provocative and provide input on the role of the interaction context in HCI and on the differences in perception and responses between human and artificial systems. Specifically, different effects observed between human and artificial systems in terms of their influence on, and reception by users, in similar social interaction scenarios, or when the same interaction characteristics, including the displayed affective profile are applied. Consequently, in a range of applications, interactive systems may demonstrate advantages, e.g., with regard to the type of responses from or interaction patterns with users, in comparison to their human counterparts. This supports their application in the role of, e.g., moderators or communication facilitators.

The experiments presented here are powerful demonstrations that Affective Dialogue Systems can be flexible participants in dyadic and multi-interactant scenarios. The findings extend previous results on conversational agents with different "personalities" (e.g., Bevacqua, de Sevin, Hyniewska, & Pelachaud, 2012) by following specific conversational goals to achieve specific social consequences. In this sense we provide evidence that such systems can be useful tools in social and behavioural research on the one hand, and on the other hand, the specific results highlight differences between responses to humans and to artificial systems that underscore the potential usefulness of employing such systems in social networks and possibly offline in embodied systems.